\documentclass[12pt,preprint]{aastex}

\shortauthors{Straughn et al.}

\usepackage{subfigure}
\usepackage{natbib}

\newcommand{\etal}	{\mbox{et al.\,}}

\newcommand{\Mo}	{\mbox{M$_{\odot}$}}

\newcommand{\cle}	{\ensuremath{\lesssim}}
\newcommand{\cge}	{\ensuremath{\gtrsim}}

\newcommand{\Hline}[1]{\mbox{H{\footnotesize {#1}}}}
\newcommand{\Ha}{\Hline{\mbox{$\alpha$}}\thinspace}
\newcommand{\Hb}{\Hline{\mbox{$\beta$}}\thinspace}

\newcommand{\OII}{[O\,{\sc ii}]}
\newcommand{\SII}{[S\,{\sc ii}]}
\newcommand{\SIII}{[S\,{\sc iii}]}
\newcommand{\OIII}{[O\,{\sc iii}\thinspace]}

\begin{document}

\title{\emph{Hubble Space Telescope} WFC3 Early Release Science: Emission--Line Galaxies from Infrared Grism Observations}

\author{
Amber N. Straughn\altaffilmark{1,2}, 
%A. N. Straughn\altaffilmark{1,2}, 
Harald Kuntschner\altaffilmark{3}, 
%H. Kuntschner\altaffilmark{3}, 
Martin K{\"u}mmel\altaffilmark{3}, 
%M. K{\"u}mmel\altaffilmark{3}, 
Jeremy R. Walsh\altaffilmark{3}, 
%J. R. Walsh\altaffilmark{3},
Seth H. Cohen\altaffilmark{4}, 
%S. H. Cohen\altaffilmark{4},  
Jonathan P. Gardner\altaffilmark{2}, 
%J.P. Gardner\altaffilmark{2}, 
Rogier A. Windhorst\altaffilmark{4},
%R. A. Windhorst\altaffilmark{4},
Robert W.~O'Connell\altaffilmark{5},
Norbert Pirzkal\altaffilmark{6}, 
Gerhardt Meurer\altaffilmark{7},
%G. Meurer\altaffilmark{7},
Patrick J.~McCarthy\altaffilmark{8}, 
%Nimish P. Hathi\altaffilmark{8}, 
Nimish P. Hathi\altaffilmark{9}, 
Sangeeta Malhotra\altaffilmark{4},
James Rhoads\altaffilmark{4},
  %start SOC below.
Bruce Balick\altaffilmark{9},
Howard E.~Bond\altaffilmark{6},
Daniela Calzetti\altaffilmark{10},
%C.~M.~Carollo\altaffilmark{11},
Michael J.~Disney\altaffilmark{11},
Michael A.~Dopita\altaffilmark{12},
Jay A.~Frogel\altaffilmark{13},
Donald N.~B.~Hall\altaffilmark{14},
Jon A.~Holtzman\altaffilmark{15},
Randy A.~Kimble\altaffilmark{2},
Max Mutchler\altaffilmark{6}
%G.~Luppino\altaffilmark{14},
%P.~J.~McCarthy\altaffilmark{11},
Francesco Paresce\altaffilmark{16},
Abhijit Saha\altaffilmark{17},
Joseph I.~Silk\altaffilmark{18},
John T.~Trauger\altaffilmark{19},
Alistair R.~Walker\altaffilmark{20},
Bradley C.~Whitmore\altaffilmark{6},
%R.~A.~Windhorst\altaffilmark{18},
Erick T.~Young\altaffilmark{21},
Chun Xu\altaffilmark{22} 
}

\altaffiltext{1}{amber.n.straughn@nasa.gov}
\altaffiltext{2}{Astrophysics Science Division, Goddard Space Flight Center, Code 665, Greenbelt, MD 20771}
\altaffiltext{3}{Space Telescope European Coordinating Facility, Karl Schwarzschild Str. 2, D 85748 Garching, Germany}
\altaffiltext{4}{School of Earth and Space Exploration, Arizona State University, Tempe, AZ 85287}
\altaffiltext{5}{Department of Astronomy, University of Virginia, Charlottesville, VA 22904-4325}
\altaffiltext{6}{Space Telescope Science Institute, Baltimore, MD 21218}
\altaffiltext{7}{International Centre for Radio Astronomy Research, The University of Western Australia, Crawley WA, 6009}
\altaffiltext{8}{Observatories of the Carnegie Institute of Washington, Pasadena, CA 91101}
\altaffiltext{9}{Department of Physics \& Astronomy, University of California, Riverside, CA 92521}

% start SOC below:
\altaffiltext{9}{Department of Astronomy, University of Washington, Seattle, WA 98195-1580}
\altaffiltext{10}{Department of Astronomy, University of Massachusetts, Amherst, MA 01003}
%\altaffiltext{11}{Department of Physics, ETH-Zurich, Zurich, 8093 Switzerland}
\altaffiltext{11}{School of Physics and Astronomy, Cardiff University, Cardiff CF24 3AA, United Kingdom}
\altaffiltext{12}{Research School of Astronomy \& Astrophysics,  The Australian National University, ACT 2611, Australia}
\altaffiltext{13}{Association of Universities for Research in Astronomy, Washington, DC 20005}
\altaffiltext{14}{Institute for Astronomy, University of Hawaii, Honolulu, HI 96822}
\altaffiltext{15}{Department of Astronomy, New Mexico State University, Las Cruces, NM 88003}
%\altaffiltext{11}{Observatories of the Carnegie Institution of Washington, Pasadena, CA 91101-1292}
\altaffiltext{16}{Istituto di Astrofisica Spaziale e Fisica Cosmica, INAF, Via Gobetti 101, 40129 Bologna, Italy}
\altaffiltext{17}{National Optical Astronomy Observatories, Tucson, AZ 85726-6732}
\altaffiltext{18}{Department of Physics, University of Oxford, Oxford OX1 3PU, United Kingdom}
\altaffiltext{19}{NASA--Jet Propulsion Laboratory, Pasadena, CA 91109}
\altaffiltext{20}{Cerro Tololo Inter-American Observatory, La Serena, Chile}
%\altaffiltext{18}{School of Earth and Space Exploration, Arizona State University, Tempe, AZ 85287-1404}
\altaffiltext{21}{NASA--Ames Research Center, Moffett Field, CA 94035}
\altaffiltext{22}{Shanghai Institute of Technical Physics, 200083 Shanghai, China}

\keywords{catalogs ---  techniques: spectroscopic ---  galaxies: starburst}

\begin{abstract}
We present grism spectra of emission--line galaxies (ELGs) from 0.6--1.6 microns from the Wide Field Camera 3 (WFC3) on the Hubble Space Telescope (HST).  These new infrared grism data augment previous optical Advanced Camera for Surveys G800L 0.6--0.95 micron grism data in GOODS--South from the PEARS program, extending the wavelength covereage well past the G800L red cutoff.  The ERS grism field was observed at a depth of 2 orbits per grism, yielding spectra of hundreds of faint objects, a subset of which are presented here.  ELGs are studied via the \Ha, \OIII, and \OII\ emission lines detected in the redshift ranges 0.2$\cle$z$\cle$1.4, 1.2$\cle$z$\cle$2.2 and 2.0$\cle$z$\cle$3.3 respectively in the G102 (0.8--1.1 microns; R$\simeq$210)  and G141  (1.1--1.6 microns; R$\simeq$130)  grisms.  The higher spectral resolution afforded by the WFC3 grisms also reveals emission lines not detectable with the G800L grism (e.g., \SII\ and \SIII\ lines).   From these relatively shallow observations, line luminosities, star--formation rates, and grism spectroscopic redshifts are determined for a total of 48 ELGs to m$_{AB(F098M)}$$\simeq$25 mag.  Seventeen GOODS--South galaxies that previously only had photometric redshifts now have new grism--spectroscopic redshifts, in some cases with large corrections to the photometric redshifts ($\Delta$z$\simeq$0.3--0.5).  Additionally, one galaxy had no previously--measured redshift but now has a secure grism--spectroscopic redshift, for a total of 18 new GOODS--South spectroscopic redshifts.  The faintest source in our sample has a magnitude m$_{AB(F098M)}$$=$26.9 mag.  The ERS grism data also reflect the expected trend of lower specific star formation rates for the highest mass galaxies in the sample as a function of redshift, consistent with downsizing and discovered previously from large surveys.  These results demonstrate the remarkable efficiency and capability of the WFC3 NIR grisms for measuring galaxy properties to faint magnitudes and redshifts to z$\cge$2.

\end{abstract}

%%%%%%%%%%%%%%%%%%%%%%%%%  INTRO: SEC. 1 %%%%%%%%%%%%%%%%%%%%%%%%%
\section{Introduction}
Galaxies that are actively star--forming make up a distinct population of sources that are involved in ongoing evolution---that is, they are in the very process of converting gas into stars and thereby changing their chemical content and stellar mass.  Star--forming galaxies are also often associated with larger scale galaxy evolution across cosmic time, in that galaxy interactions are often found to cause enhanced star formation (e.g., Li \etal 2008, Overzier \etal 2008, Larson \& Tinsley 1978) and galaxy evolution as a whole is thought to proceed hierarchically via galaxy interactions and merging (e.g., White \& Frenk 1991; Navarro \etal 1997, etc.).  These actively star--forming galaxies are therefore important to study within the overall context of galaxy assembly.  Information about star formation activity is revealed in the galaxies' emission lines, particularly \Ha, \OIII, and \OII\ at rest--frame wavelengths $\lambda$6563{\AA}, $\lambda$$\lambda$4959{\AA}, 5007{\AA}, and $\lambda$3727{\AA} respectively.  Many studies have used emission lines to investigate the star--forming properties of galaxies over various redshift ranges (Hammer \etal 1997; Kennicutt 1983; Gallego \etal 1995; Kewley \etal 2004; NICMOS grism study: McCarthy \etal 1999, Yan \etal 1999; WISP grism Survey: Atek \etal 2010).

The installation of the new Wide Field Camera 3 (WFC3) on the Hubble Space Telescope (HST) in mid--2009 has provided a new capability for studying star formation and has already resulted in a variety of scientific discoveries in observational cosmology.  Particularly, the increase in sensitivity, field of view, and resolution of the WFC3/IR over previous infrared instrumentation has been used to detect some candidates for the most distant galaxies ever observed (Bouwens \etal 2010, Yan \etal 2010a, Oesch \etal 2010, McLure \etal 2010, Finkelstein \etal 2010) etc.) from the ultra--deep WFC3 imaging (Illingworth \etal PID GO--11563) of the Hubble Ultra Deep Field (HUDF; Beckwith \etal 2006).  The less deep but wider--area broadband data from the Early Release Science (ERS) II (PI O'Connell, PID GO--11359) program have also been used to study high--redshift candidates (Wilkins \etal 2010, Labbe \etal 2010, Bouwens \etal 2010, Yan \etal 2010b) and UV--dropout galaxies (Hathi \etal 2010).  Windhorst \etal (2010)  describe the WFC3 ERS program in detail, which we summarize in Section 2.

In addition to the broadband data used for most of these studies, the ERS II program also consists of one field observed with both the G102 (0.8--1.1 microns; R$\simeq$210)  and G141 (1.1--1.6 microns; R$\simeq$130) infrared grisms (described in detail below).  van Dokkum \etal (2010) report on a bright z$=$1.9 compact galaxy in the ERS grism data.  Here, we present emission--line galaxies from the WFC3 ERS grism observations, demonstrating the unique capability of this instrument for detecting star--forming galaxies in the infrared reaching to magnitudes m$_{AB(F098M)}$$\simeq$25 mag with only 2 orbits of HST time.  By searching for emission lines in the infrared grism data, we are able to push detection of these galaxies and subsequent measurement of their physical properties to redshifts z$\simeq$2.0.  Grism studies with HST's Advanced Camera for Surveys (ACS) G800L grism have proven successful at detecting emission--line galaxies (ELGs) in the optical (Meurer \etal 2007, Straughn \etal 2008, Straughn \etal 2009, Xu \etal 2007, Pirzkal \etal 2006), and here we extend these studies of ELGs to the infrared.

%The ACS grism has also been used to study faint z$\simeq$5 Lyman break galaxies (Rhoads \etal 2009), early--type galaxies (Ferreras \etal 2009), AGN (Grogin \etal 2007), z$\simeq$6 galaxies (Malhotra \etal 2005), and galaxy clusters (Tsvetanov \etal 2002).

%%%%%%%%%%%%%%%%%%%%%%%%%  DATA: SEC. 2 %%%%%%%%%%%%%%%%%%%%%%%%%
\section{Data}
The ERS II program for WFC3 consists of both UV and IR observations of about 30\% of the GOODS--South field (Giavalisco \etal 2004).  Here we summarize the ERS II program; Windhorst \etal (2010) present the ERS II data reduction effort in detail.  Eight pointings were imaged with the UVIS channel (UV filters F225W, F275W, and F336W at depths of 2 orbits/pointing/filter for F225W and F275W and 1 orbit/pointing for F336W) and ten with the IR channel (filters F098M, F125W, F160W) at 2 orbits/pointing/filter.  Grism observations of one WFC3 pointing (c.f. Fig. 1) were performed using the infrared ``blue'' G102 grism (R$\simeq$ 210)  and the ``red'' (R$\simeq$130) G141 grism, providing spectral coverage from 0.8--1.6 microns at 2 orbits/grism depth.  

The WFC3 IR channel has a field of view of 4.65 arcminutes$^{2}$ at a resolution of 0.19 arcsec/pixel.  The ten 5 x 2 grid pattern WFC3/ERS II IR pointings span the northern $\simeq$40 square arcminutes of GOODS--South, providing new high--resolution infrared imaging to accompany this widely--used multiwavelength dataset.  The ERS II grism field lies in the north--central region of the ERS II imaging field (Fig. 1; J2000 53.071121 -27.709646) and is overlapped completely by the HST Probing Evolution And Reionization Spectroscopically (PEARS; Malhotra PID 10530) ACS grism survey South Field $\#$4.  As such, combined with the earlier ACS data, these new infrared grism data provide unprecedented spectral grism coverage in the optical to infrared wavelength range of objects reaching to continuum magnitude m$_{AB(F098M)}$$\simeq$25 mag.

%%%%%%%%%%%%%%%%%%%%%%%%%  ANALYSIS: SEC. 3 %%%%%%%%%%%%%%%%%%%%%%%%%
\section{Analysis}
%Here we summarize the procedures used to extract slitless spectra and measure physical properties of ELGs in the WFC3 ERS II grism field, and refer the reader to Straughn \etal (2008) for a full description of this methodology employed with the PEARS data (see also Meurer \etal 2007 for a general discussion of processing ACS grism data).

The latest version (March 11, 2010) STSDAS \emph{CALWF3} v2.0 pipeline reduced direct imaging and the associated grism exposures were
obtained from the MAST Archive.   The direct images were combined for each of the F098M and F140W filters using Multidrizzle (Koekemoer \etal 2002) and small shifts of up to 0.7 pixels were applied to individual images for alignment.  Source catalogues
were produced for each filter using SExtractor (Bertin \& Arnouts 1996). The resulting
catalogues were cleaned to remove spurious sources at the edges and artifacts from persistence
effects caused by a bright star in the preceding ERS grism observations. 

The final, cleaned source catalogues were used with the aXe grism reduction software
(version 2.0; K{\"u}mmel et al. 2009) to extract a calibrated 2--dimensional, co--added grism spectrum
for each source.  Master sky backgrounds for each grism were constructed from all publicly
available grism data as of February 2010 and subtracted from the ERS observations prior to
spectral reduction. The final 2--dimensional grism spectra include information about the associated errors and
contamination by spectra of neighboring objects. The  trace
and wavelength calibration used by aXe to extract the spectra was based on the first in--orbit
calibration observations (Kuntschner et al. 2009a,b).

%description of 1-D extraction & measurement of line fluxes etc.

\subsection{Source Selection}
We visually examined all emission--line candidate sources located in the ERS grism field; many ELGs with prominent emission lines detected in PEARS with the G800L grism also have lines in the infrared.  In addition to previously--detected PEARS ELGs, there are also sources in the field in which the strongest emission lines lie exclusively in the infrared and so were not detected in PEARS.  We used standard Gaussian fitting techniques to measure emission line fluxes and calculate SFRs based on the line luminosities, as described in the following Section.  ELGs that have line flux measurements with S/N$\ge$2 are retained in the final catalog (Table 1); 83\% of lines have S/N$\ge$3.  Whereas the PEARS pre--selected galaxies by definition have more than one line---and therefore line identifications based on the line wavelength ratio are unambiguous and in some cases already determined by the ACS grism spectra---a small number of ERS II ELG candidates only have one emission line.  For these sources, spectroscopic and photometric redshift catalogs (Grazian \etal 2006; Wuyts \etal 2008; Balestra \etal 2010; Vanzella \etal 2008) are consulted in order to determine if the source has a previously--measured redshift.  If it does, line identification is accomplished via this redshift and a grism--spectroscopic redshift (Cohen \etal 2010, in prep.; Xia \etal 2010) is calculated based on the line identification.  If it does not, the line remains unidentified (see Table 1).  However, 60\% of the ELGs have two or more emission lines, and therefore it is straightforward to assign line identifications based on the emission line wavelength ratio.  The fraction of objects with two lines in the spectra is considerably higher than in the PEARS studies (where the fraction of sources with two lines was $\simeq$30\%) for two main reasons.  First, the wavelength range for both the G102 and G141 is longer by a factor of more than 2; and second, the higher spectral resolution afforded by both WFC3 grisms allows detection of lines not previously seen in the G800L observations; namely, \SII\ at $\lambda\lambda$6716$+$6731\AA\ is now sufficiently resolved from \Ha\ (Figure 2) and \SIII at $\lambda\lambda$9069, 9532\AA\ is detected in several sources as well.  Additionally, the higher resolution allows in some cases \Hb\ and the two \OIII\ lines at $\lambda$4959 and $\lambda$5007 to be detected (\OIII\ at $\lambda\lambda$4959, 5007 can be marginally resolved in some cases; see Figure 3).  This higher resolution allows, e.g., a line identification in the extremely faint (continuum) Object 397, which has no other significant line detections in the spectrum.  Four objects in the sample are also X--ray sources and likely AGN (Szokoly \etal 2004); these are noted in Table 1.  

%The average continuum magnitude of these ACS$+$WFC3 ELGs is $AB (F098M)=$22.9 mag.

%%%%%%%%%%%%%%%%%%%%%%%%%  RESULTS: SEC. 4 %%%%%%%%%%%%%%%%%%%%%%%%%
\section{Results}
\subsection{ERS II Emission--Line Galaxy Sample}
Our final catalog of WFC3 ERS II ELGs contains 48 galaxies with a total of 73 emission lines.  Of these, 29 are \Ha, 27 \OIII, 6 \OII, 2 \SII, 2 \SIII$\lambda$9069, 2 \SIII$\lambda$9532, and 5 unidentified lines (see Table 1).  The average redshift of these galaxies is z$=$1.200 with a redshift range of z$=$0.227--2.315 (Figure 4).  The galaxies' broadband F098M magnitudes span m$_{AB(F098M)}$$=$18.67--26.87 mag with an average magnitude m$_{AB(F098M)}$$=$23.65 mag.  The faintest continuum magnitude source is Object 103 (m$_{AB(F098M)}$$=$26.87 mag) at redshift z$=$1.680 with both \OII\ and \OIII\ detected.

The ERS grism pointing lies completely within PEARS--South Field 4, and a total of 25 PEARS--detected ELGs fall into this field.  Thirteen of these have emission lines in the G102/G141 bandpasses with fluxes meeting the S/N$\cge$2 requirement (see Figure 2 for example spectra).  Many of these objects are sources with \OIII\ emission in the optical, and \Ha\ falling in the G102 grism bandpass.  One such example is Object $\#$  370  (PEARS Object $\#$ 119489; Straughn \etal 2009).  This object is also a CDF--S X--ray source (Szokoly \etal 2004; Grogin \etal 2007), which was observed to have one strong line in the optical.  We now detect another strong line in the infrared, making line identifications possible via the wavelength ratio (Figure 2).  Several PEARS--detected galaxies have \OIII\ emission near the red edge of the G800L bandpass, which overlaps with the G102 grism, and so \OIII\ is observed in G102 as well.  For these sources, \Ha\ falls into the lower--resolution G141 bandpass.   Several other single--line PEARS ELGs also have emission lines detected in G102 and/or G141 and now have grism--spectroscopic redshifts, thus demonstrating the utility of the extended wavelength/redshift range compared to ACS G800L for identifying emission lines.  Due to the higher resolution of the WFC3 grisms, the \OIII\ $\lambda$$\lambda$4959{\AA}, 5007{\AA} lines are individually detectable (though not fully resolved) in these data (e.g., Fig. 3 Objects 402 and 397), whereas they are blended in the ACS G800L grism data.  Also of note in this sample are the sources which have prominent emission lines in all three grism bandpasses---and thus different star--formation indicators.

In addition to PEARS--selected ELGs, other emission line sources were identified in the ERS II grism field that do not have lines detected in the optical G800L grism data.  Thirty--five such sources that make the S/N$\cge$2 cut are listed in Table 1.  Prevalent among these objects are ELGs with \Ha\ in the G102 bandpass; many of these also have \SIII\ detections in G141 (Fig. 2).  A few of the WFC3--detected sources are z$\simeq$2 galaxies with \OII\ and/or \OIII\ lines visible in the IR grism bandpasses (e.g., Objects 242 and 578 in Table 1).  Grism redshifts (Figure 4) are derived for all sources as described in Section 3.1 and are also presented in Table 1.  Seventeen GOODS--South galaxies that previously only had photometric redshifts now have new grism--spectroscopic redshifts.  Some of these objects have large corrections to the photometric redshifts ($\Delta$z$\simeq$0.3--0.5).  Additionally, one galaxy (Object 226) had no previously--measured redshift but now has a secure grism--spectroscopic redshift, for a total of 18 new spectroscopic redshifts.  Two sources in our catalog (402 and 474; see Figure 3 for example)
have published FORS2 and VIMOS spectroscopic redshifts (Vanzella \etal 2008; Balestra \etal 2010) that disagree with our
measurements which are listed in Table 1.

\subsection{Star Formation in ERS II ELGs}
The longer wavelength range over which to detect emission lines provides more sources with multiple lines that can be used in calculating SFRs.  In particular, \Ha---the emission line which yields the most direct and secure SFR estimate (e.g., Kennicutt 1998)---is now observed in 29 sources.  Given the low resolution of the grism spectra, some contribution from [NII] $\lambda$$\lambda$6548,6584 will be present in the measured \Ha\ line fluxes due to the blending of the lines.  The strength of this contribution in general galaxy samples varies from a few percent to factors of 0.3--0.5 for particularly massive and metal--rich galaxies (Jansen \etal 2000, Gallego \etal 1997, Kennicutt \etal 1992).  Because of this wide variation owing to differences in effective temperature, ionization, and metallicity, we do not adopt a global [NII] correction and instead note that the \Ha\ fluxes presented here are likely overestimates.  For each of the ELGs, we calculate SFRs via the prescription of Kennicutt 1998 for \Ha\ and \OII.  Using the \OIII\ line to arrive at a SFR is less secure due to the effects of metallicity and gas temperature (Kennicutt \etal 2000, Kennicutt 1992).  For eight of the ELGs in the sample, \OIII\ is the only line measured in the spectrum, and thus we use the \OIII\ SFR calibration from Straughn \etal (2009), which is derived from ELGs with both \Ha\ and \OIII\ in their spectra, and should be considered a lower limit.  Figure 5 shows the SFR as a function of redshift for these galaxies, using \Ha\ when available; then \OII\ and \OIII\, in order of preference.  The detection limit is evident in this plot; we see the expected bias toward higher SFRs at higher redshifts.  

%Galaxy stellar masses were calculated for the sample via spectral energy distribution (SED) fits to the 10--band (WFC3 $+$ ACS) photometry (Cohen \etal 2010, in preparation).  
The broad--band spectral energy distributions (SEDs), using ten band WFC3 and ACS data from 0.2-1.7 microns, were fit
to a grid of models using the derived redshifts listed in Table 1, yielding galaxy stellar masses (Cohen \etal, in preparation). These Bruzual 
\& Charlot (2003) models were generated assuming an exponentially declining
star formation history, and varying the time scale $\tau$, internal extinction A$_{V}$, and age.   The specific star--formation rates (sSFRs; line--calculated SFR per unit stellar mass) of these galaxies are shown as a function of stellar mass in Figure 6, which is consistent with the general negative trend observed previously for galaxies at redshifts up to z$\simeq$2 (e.g., Feulner \etal 2005, Bauer \etal 2005, Erb \etal 2006, Elbaz \etal 2007, Noeske \etal 2007, Rodighiero \etal 2010, etc.).  Since our sample is selected based on the presence
of emission lines (and thus ongoing star formation) we do not expect
to see the galaxies with much lower SFRs which would likely occupy the
lower-left area of Fig. 6 (detection limits also affect the
selection of low--SFR galaxies).  While the sample is clearly small due
to the limited volume observed and we would likely miss any rare,
high--mass, high--sSFR sources that might occupy the upper--right area of
Figure 6, previous studies using large galaxy samples including
Spitzer IRAC 3.6 micron data indicate that these sources are in fact not
missed in these studies---i.e., that the upper limit to the galaxies'
SFR as a function of mass is real (Noeske \etal 2007).  

In Figure 7 we show the sSFR as a function of redshift in different mass bins, which is consistent with previous studies showing the trend of lower sSFR for higher mass galaxies as a function of redshift (e.g., Zheng \etal 2007, Damen \etal 2009, Rodighiero \etal 2010).  While these previous studies investigating the sSFR as a function of redshift have made use of very large samples of galaxies from an array of large observational programs (COMBO--17, Zheng \etal 2007; SIMPLE---Spitzer IRAC and MUSYC, Fazio \etal 2004, Damen \etal 2009; GOODS, Dickinson \etal 2003; see also Martin \etal 2007 and Perez--Gonzalez \etal 2008), we have demonstrated here that the WFC3 IR grism data generally reflect these trends using only two orbits of HST time.  Future observations of this type will serve to investigate these preliminary trends in a more statistically significant way.

%%%%%%%%%%%%%%%%%%%%%%%%%  SUMMARY: SEC. 5 %%%%%%%%%%%%%%%%%%%%%%%%%
\section{Summary}
We detect a total of 73 emission lines from 48 galaxies in the ERS II grism field, allowing calculation of line fluxes, SFRs, and grism--spectroscopic redshifts.  Thirteen of these galaxies had emission lines in the optical ACS G800L grism data and 35 are newly-detected star--forming galaxies with emission lines in the infrared.  We show the SFRs of these galaxies as a function of redshift and discuss trends involving SFRs and galaxy masses and redshifts.  These data are consistent with previous studies showing that the sSFRs of the most massive (M$>$10$^{11}$$\Mo$) star--forming galaxies are generally lower than their lower mass counterparts as a function of redshift (Zheng \etal 2007, Damen \etal 2009).  These data demonstrate the efficiency of the WFC3 grisms in detecting faint star--forming galaxies at z$\simeq$0.2--2.5.  This work sets the stage for larger area and deeper studies of star--forming galaxies with WFC3 in the future, which will serve to greatly increase the sample size and statistics, and will probe even fainter and less massive sources. 

This research was supported in part by an appointment to the NASA Postdoctoral Program at Goddard Space Flight Center, administered by Oak Ridge Associated Universities through a contract with NASA (ANS).  This paper is based on Early Release Science observations made by the WFC3 Scientific Oversight Committee.  We thank the anonymous referee for suggestions which improved the paper.   We are grateful to the Director of the Space Telescope Science Institute for awarding Director's Discretionary time for this program.  Finally, we are deeply indebted to the brave astronauts of STS-125 for rejuvenating HST.

%%%%%%%%%%%%  FIGURE 1: ERS LAYOUT %%%%%%%%%%%%%%%%%%
\begin{figure}
\includegraphics[scale=0.6]{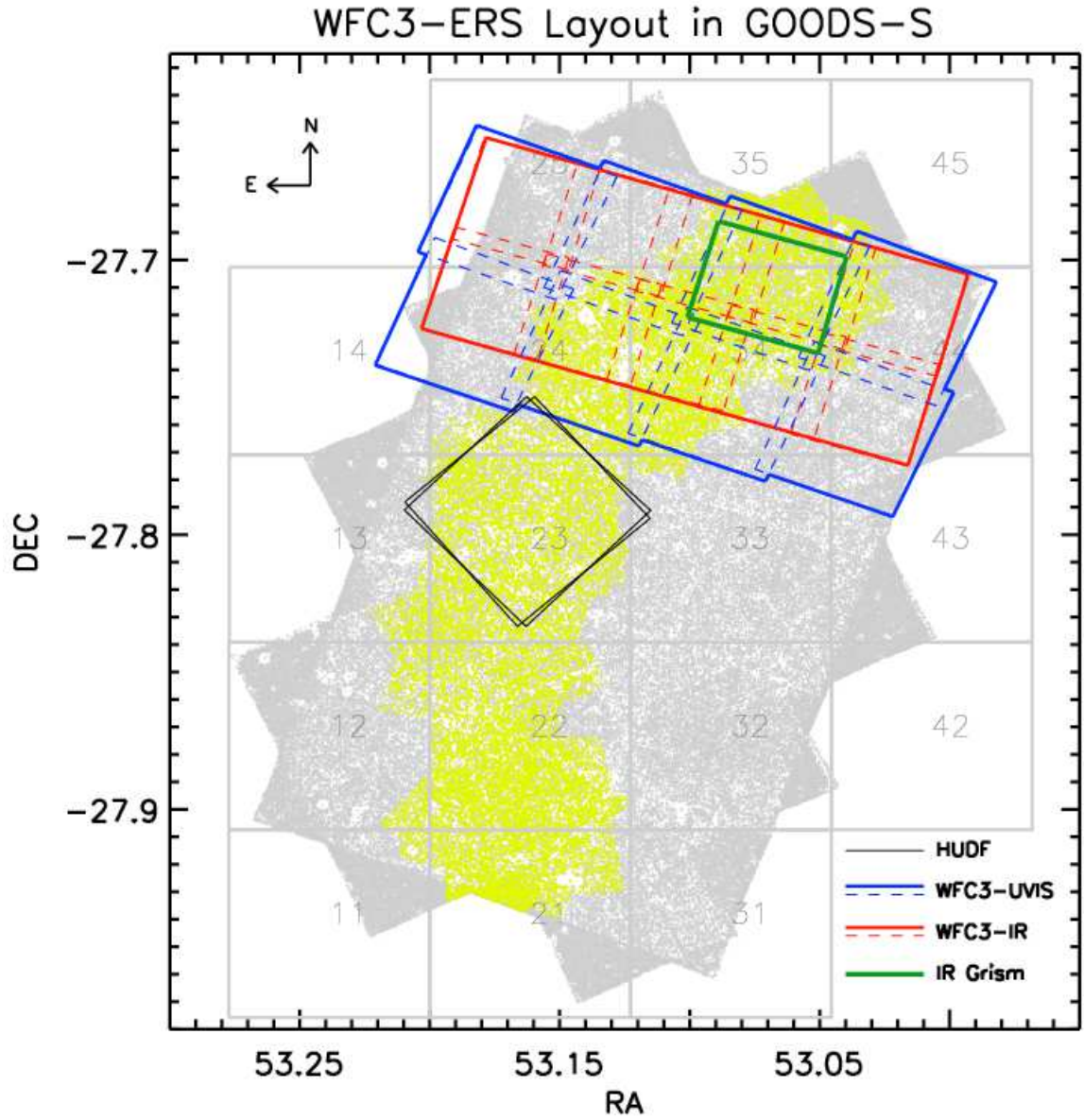}
\caption{ERS II Layout in GOODS--South.  Grey area is ACS GOODS--South with tile numbers for that dataset.  Yellow areas are the five ACS PEARS grism fields; the black box is ACS HUDF.  The WFC3 ERS II UVIS fields are outlined in blue, and WFC3 ERS II IR fields are outlined in red.  The green box is the WFC3 ERS II IR grism field. }
\label{fig:ewdist}
\end{figure}

%%%%%%%%%%%%  FIGURE 2: GRISM THROUGHPUT  %%%%%%%%%%%%%%%%%%
%\begin{figure}
%\centering
%\includegraphics[scale=0.75]{grismthrougput.ps}
%\caption{Total throughput for the G102 (blue) and G141 (red) IR grism modes.}
%\label{fig:ewdist}
%\end{figure}

%%%%%%%%%%%%  FIGURE 2: PEARS PRE-SEL ELGS  %%%%%%%%%%%%%%%%%%
\begin{figure}
\centering
\subfigure{
\includegraphics[scale=0.5]{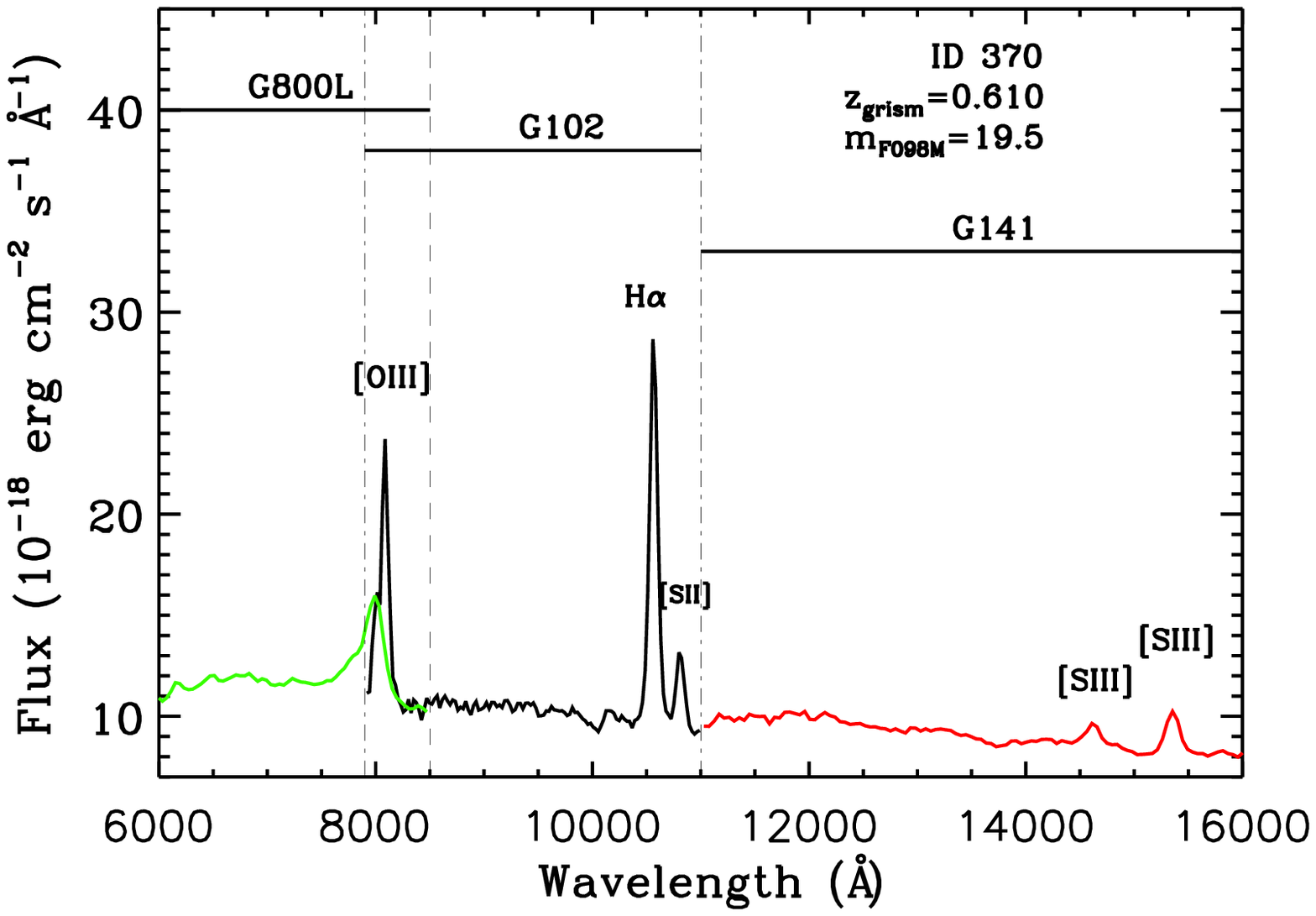}}
\hspace{0cm}
\subfigure{
\includegraphics[scale=0.5]{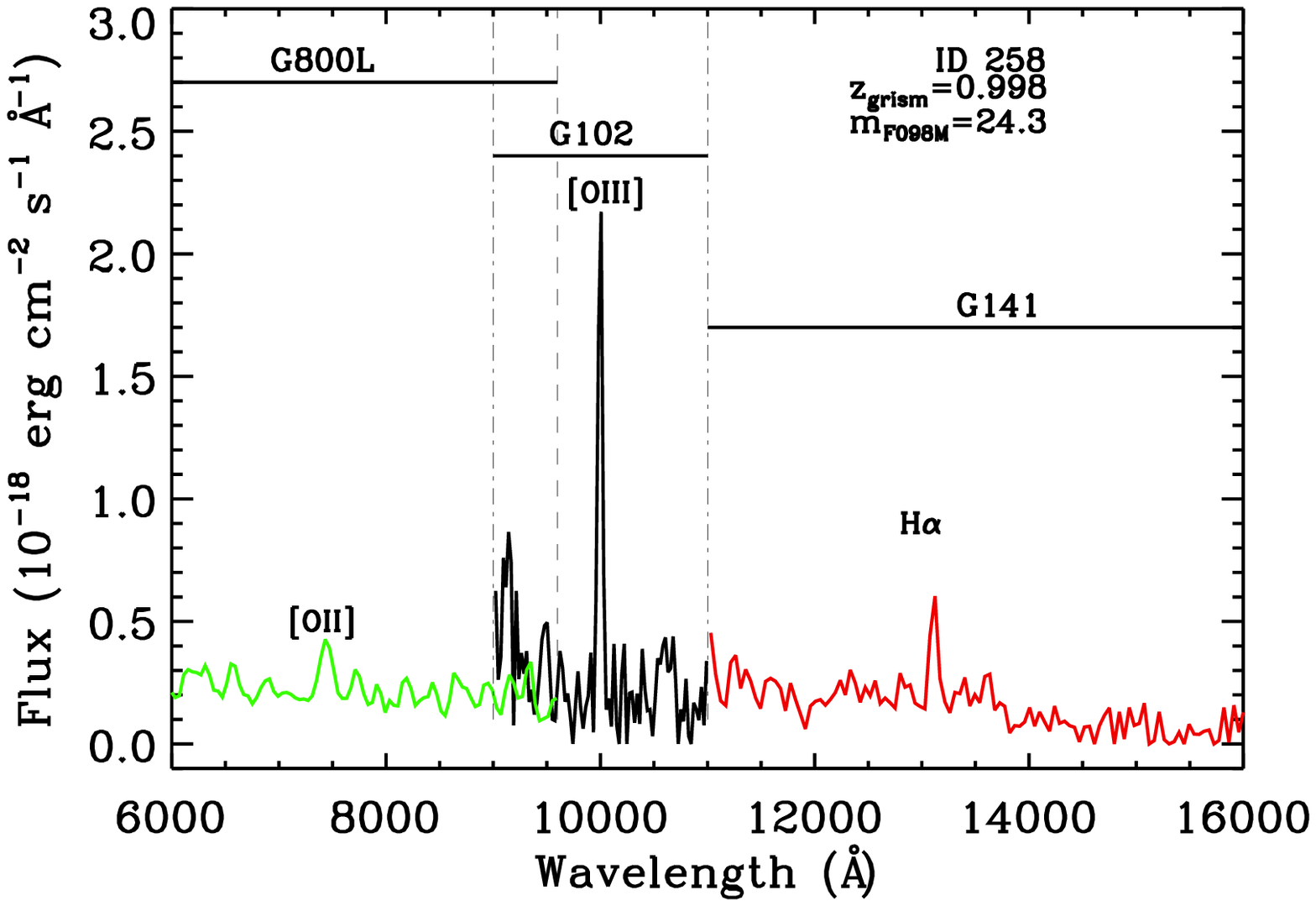}}
\hspace{0cm}
\subfigure{
\includegraphics[scale=0.5]{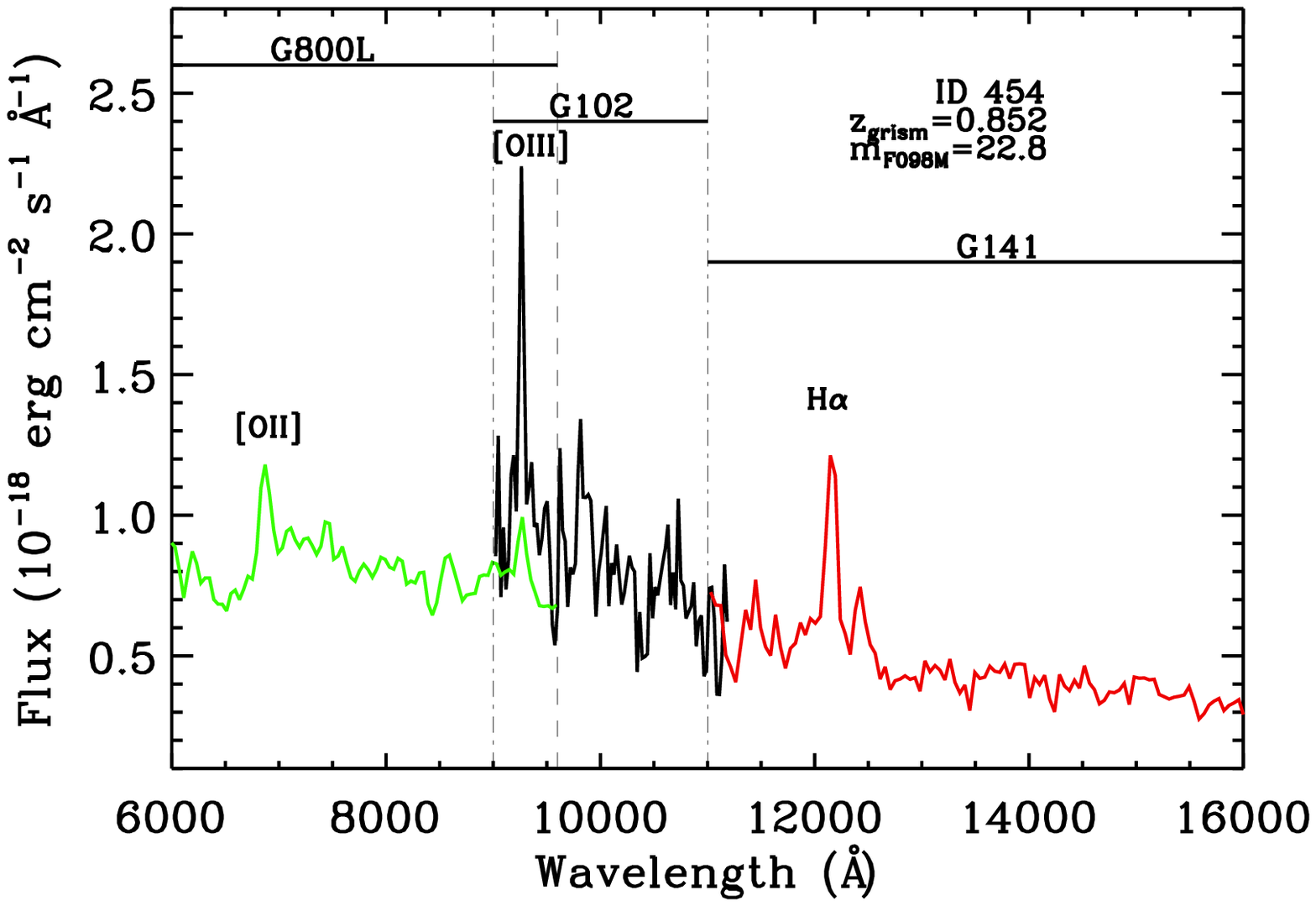}}
\caption{Three example grism spectra for ELGs (Table 1 Objects 370, 258, 454 from top to bottom) pre--selected from the ACS PEARS grism ELG study of Straughn \etal (2009).  The ACS G800L data are shown in green; WFC3 G102 in black, and WFC3 G141 in red.  The ACS G800L fluxes have been scaled for visual purposes to match the WFC3 data.  Compared to ACS G800L, the higher--resolution WFC3 grisms allow detection of the [S II] and [S III] lines.}
\label{fig:compspec}
\end{figure}

%%%%%%%%%%%%  FIGURE 3: WFC3 NEW ELGS  %%%%%%%%%%%%%%%%%%
\begin{figure}
\centering
\subfigure{
\includegraphics[scale=0.5]{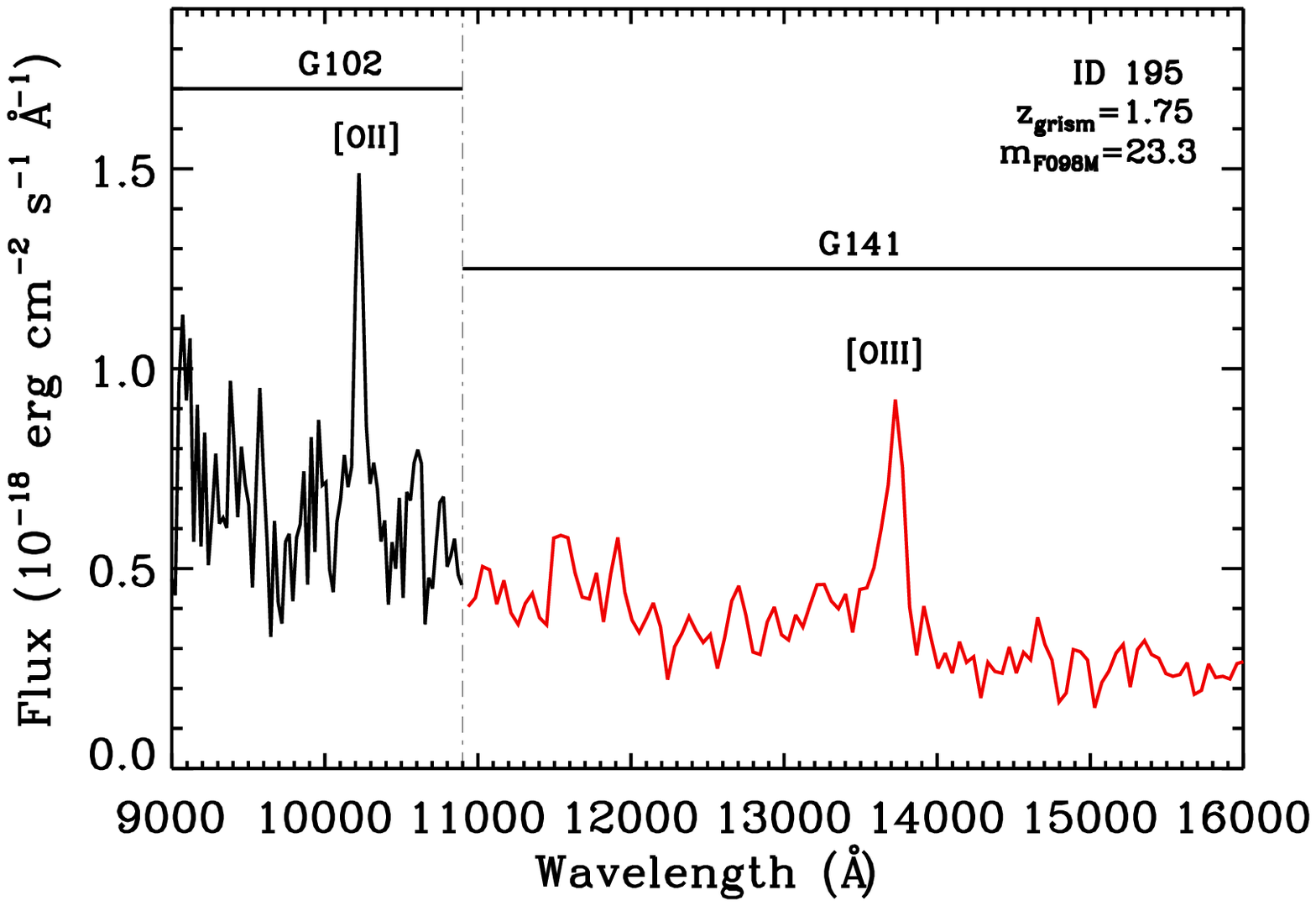}}
\hspace{0cm}
\subfigure{
\includegraphics[scale=0.5]{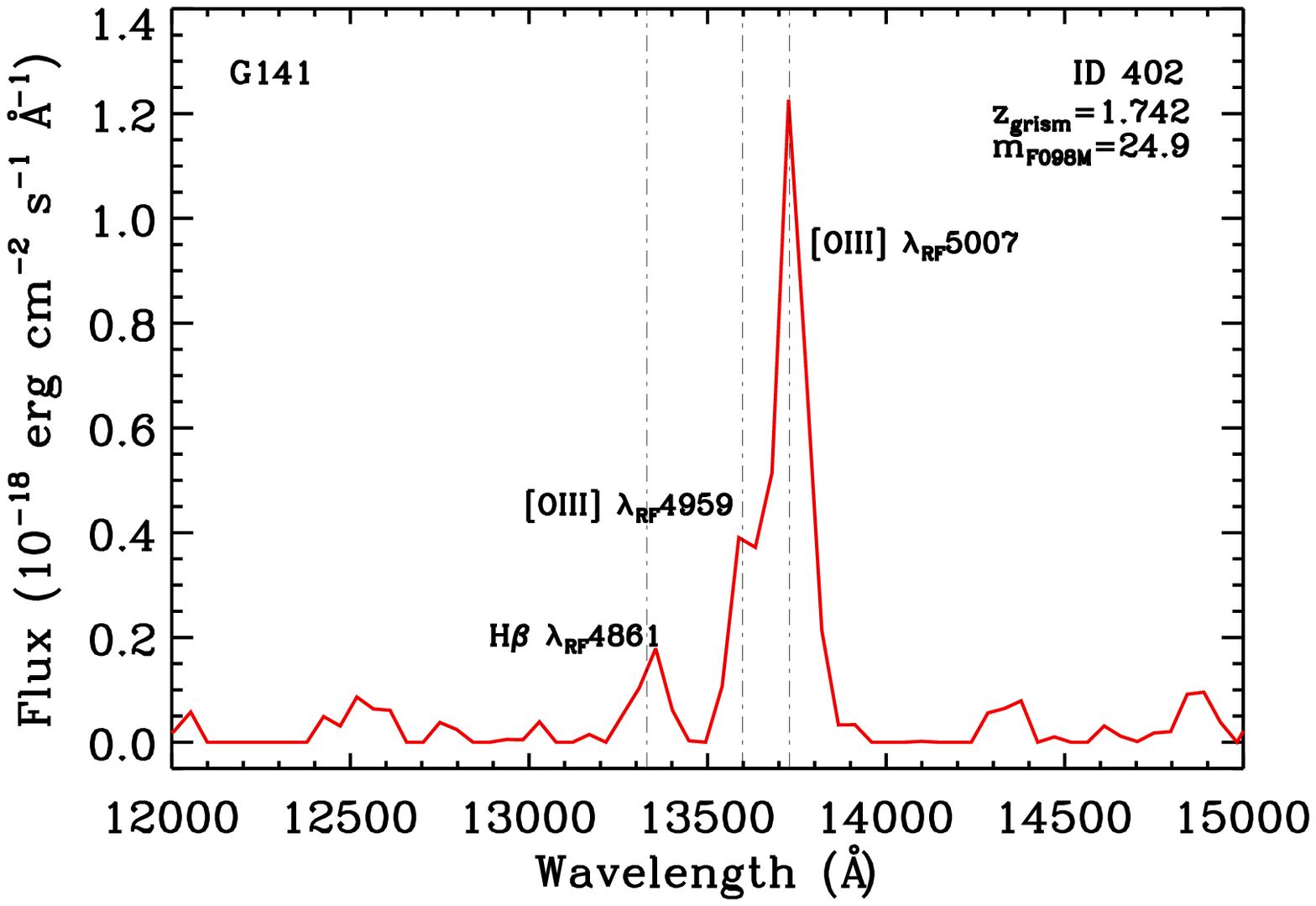}}
\hspace{0cm}
\subfigure{
\includegraphics[scale=0.5]{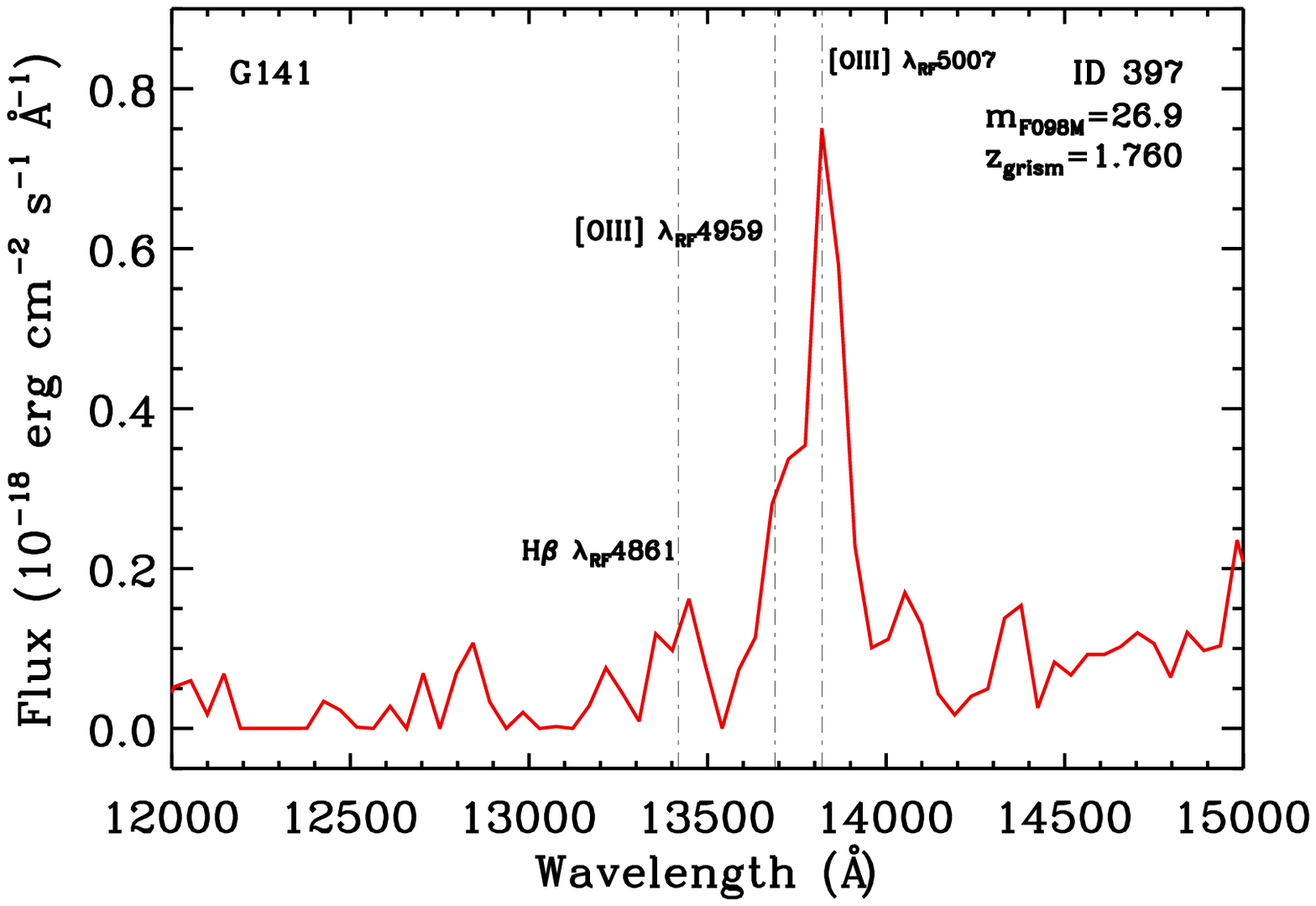}}
\caption{Three example grism spectra for ELGs with detected emission lines only in the IR grisms.  Of note in the sample are intermediate redshift galaxies with extremely faint continuum fluxes and strong emission lines, as exemplified by Objects 402 and 397 here.  The higher resolution of the WFC3 grisms allows detection of the two \OIII\ $\lambda$$\lambda$4959{\AA}, 5007{\AA} lines (though not resolved) along with \Hb, providing a line identification for these faint sources with no previously--measured redshifts.}
\label{fig:compspec}
\end{figure}

\clearpage

%%%%%%%%%%%%  FIGURE 4: ZDIST %%%%%%%%%%%%%%%%%%
\begin{figure}
\centering
\includegraphics[scale=0.75]{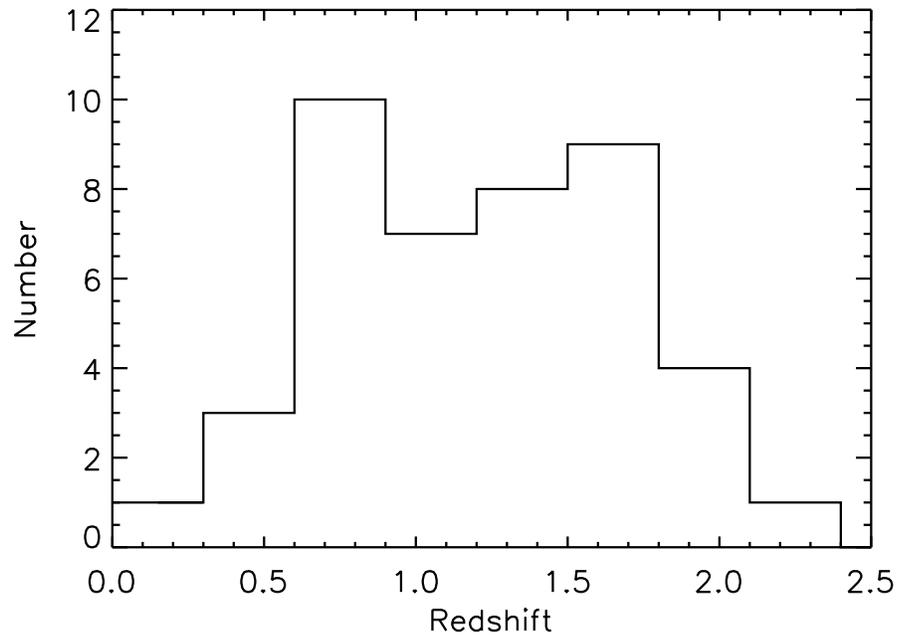}
\caption{Grism--spectroscopic redshift distribution of WFC3 ELGs.  The \Ha, \OIII, and \OII\ emission lines are visible in the WFC3 grism bandpasses at redshifts 0.2$\cle$z$\cle$1.4, 1.2$\cle$z$\cle$2.2 and 2.0$\cle$z$\cle$3.3 respectively.  The majority of galaxies have more than one emission line, allowing secure line identifications and grism redshift determination. }
%\label{fig:compspec}
\end{figure}

%%%%%%%%%%%%  FIGURE 5: sSFR VS. M  %%%%%%%%%%%%%%%%%%
\begin{figure}
\centering
\includegraphics[scale=0.75]{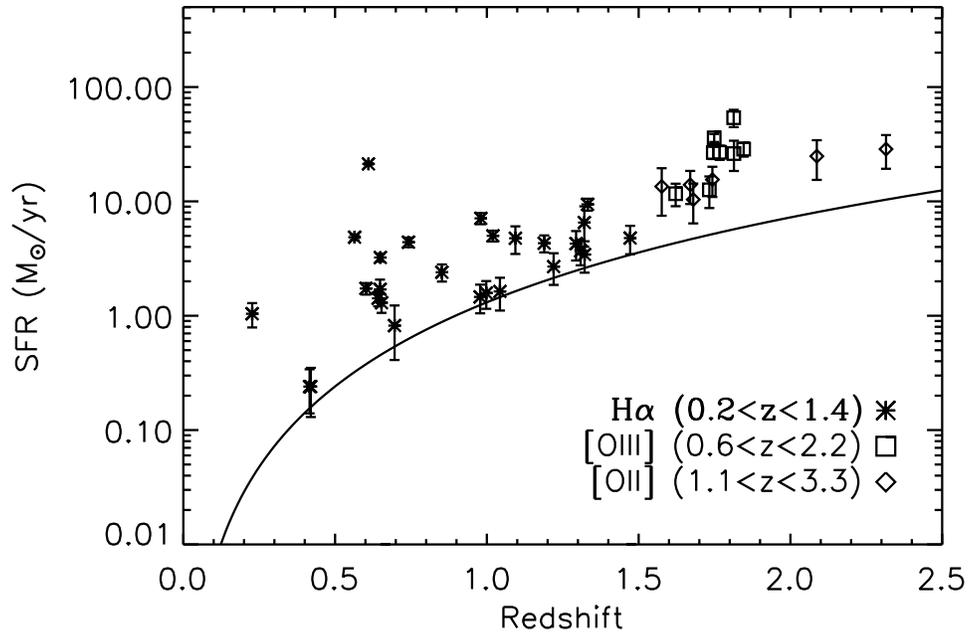}
\caption{Star--formation rate for grism--detected sources as a function of redshift.  SFRs are calculated from \Ha, \OII, and \OIII\ line fluxes in order of preference, as described in Section 4.3.}
%\label{fig:compspec}
\end{figure}

%\begin{figure}
%\centering
%\subfigure{
%\includegraphics[scale=0.75]{/Users/astraughn/WFC3/SFR_comps_xy.ps}}
%\hspace{0cm}
%\subfigure{
%\includegraphics[scale=0.75]{/Users/astraughn/WFC3/SFR_comps_abs.ps}}
%\caption{caption here...WFC3 ERS II ELGs.}
%\label{fig:compspec}
%\end{figure}

%%%%%%%%%%%%  FIGURE 8: SFR COMPS: EL & SED  %%%%%%%%%%%%%%%%%%
%\begin{figure}
%\centering
%\includegraphics[scale=0.75]{/Users/astraughn/WFC3/SFR_compsv2.ps}
%\hspace{0cm}
%\includegraphics[scale=0.75]{/Users/astraughn/WFC3/SFR_comps_z.ps}}
%\caption{Comparison of SFRs calculated from both the emission--line fluxes and the SED fits. }
%\label{fig:compspec}
%\end{figure}

%%%%%%%%%%%%  FIGURE 6: SSFR AS FCN. OF M %%%%%%%%%%%%%%%%%%
\begin{figure}
\centering
\includegraphics[scale=0.75]{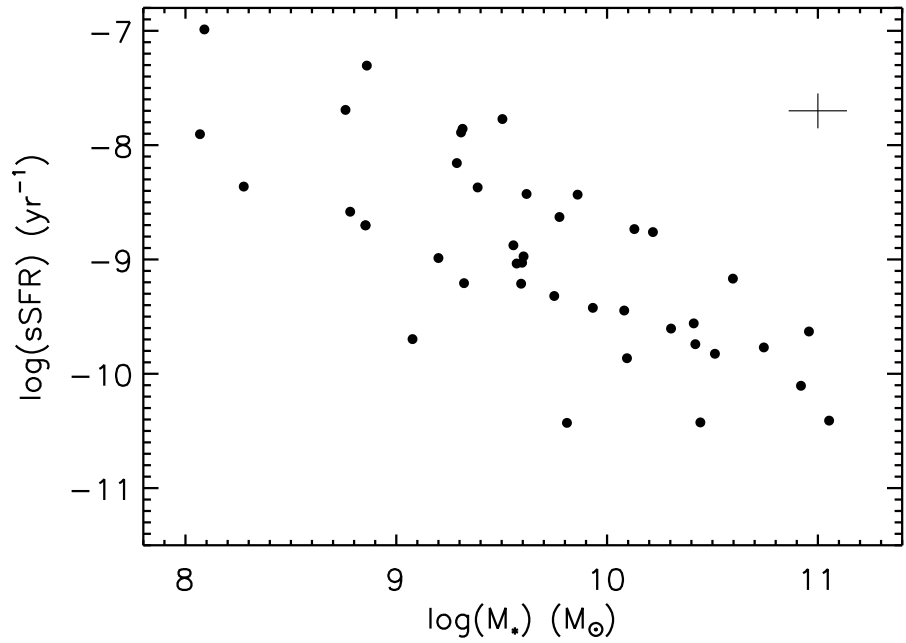}
\caption{Specific star--formation rate as a function of stellar mass.  The SFRs are computed by line flux measurements and the stellar masses are calculated from SED fits (Section 4.3).  The results from the WFC3 IR grism data are consistent with previous studies showing the relation of lower sSFR at higher mass.  Average error estimate for the sample is shown in the upper--right.}
%\label{fig:compspec}
\end{figure}

%%%%%%%%%%%%  FIGURE 7: SSFR AS FCN. OF Z %%%%%%%%%%%%%%%%%%
\begin{figure}
\centering
\includegraphics[scale=0.75]{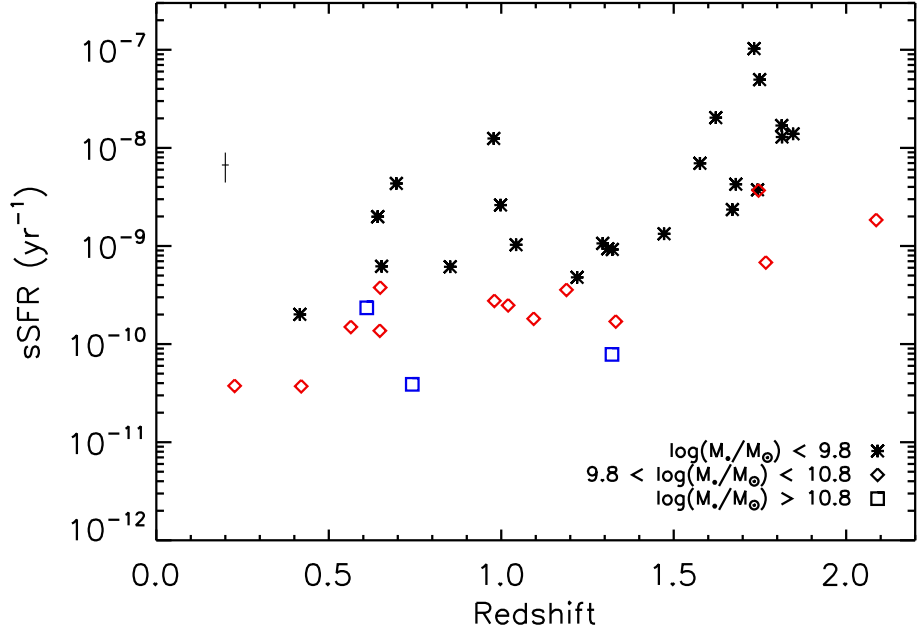}
\caption{Specific star--formation rate as a function of redshift for three different mass bins.  Although the sample size is small, the 2--orbit WFC3 grism data results are consistent with previous large survey studies that show the general trend of lower sSFRs for the highest mass galaxies as a function of redshift.  Average error estimations are shown in upper--left.  The lowest redshift source in the highest mass bin (blue squares) is an AGN, so some contribution to the line flux from the AGN is expected and likely results in a higher calculated SFR (uncorrected for AGN contamination).}
%\label{fig:compspec}
\end{figure}

%%%%%%%%%%%%  TABLE 1  %%%%%%%%%%%%%%%%%%
\begin{deluxetable}{ccccccccccl}
\rotate
\tabletypesize{\scriptsize}
\tablecaption{Global Properties of Emission--Line Galaxies \label{table1}}
\tablewidth{0pt}
\tablehead{ \colhead{ID}&\colhead{RA}&\colhead{Dec}&\colhead{$AB (F098M)$}&\colhead{Wavelength}&\colhead{Flux}&\colhead{EW}&\colhead{Line}&\colhead{SFR}&\colhead{Grism}&\colhead{Flag}\\
\colhead{}&\colhead{(deg)}&\colhead{(deg)}&\colhead{(mag)}&\colhead{(\AA)}&\colhead{($10^{-18} erg/s/cm^2$)}&\colhead{(\AA)}&\colhead{ID}&\colhead{\Mo\ yr$^{-1}$}&\colhead{Redshift}&\colhead{--}
}
\startdata
432 & 53.0484200 & -27.7095337 & 24.45 & 12850 & 94.7$\pm$18 & 114 & \OIII\ & 19.7$\pm$3 & 1.573 & 1 \\
& & & & 9599 & 60.0$\pm$26 & 35 & \OII\ & 13.5$\pm$6 & 1.573 & 1 \\
499 & 53.0515862 & -27.7047272 & 25.06 & 16226 & 44.5$\pm$12 & 199 & \Ha\ & 4.8$\pm$1 & 1.472 & 1 \\
& & & & 12381 & 37.3$\pm$16 & 176 & \OIII\ & 6.7$\pm$2 & 1.472 & 1 \\
370\tablenotemark{\dag}  & 53.0551300 & -27.7113667 & 19.51 & 8080 & 1140.5$\pm$91 & 62 & \OIII\ & 23.7$\pm$1 & 0.610 & 1 \\
& & & & 14620 & 166.6$\pm$16 & 22 & \SIII $\lambda$9069 & -- & -- & 1 \\
& & & & 10810 & 344.9$\pm$22 & 35 & \SII & -- & -- & 1 \\
& & & & 10565 & 1733.6$\pm$23 & 98 & \Ha\ & 21.2$\pm$0 & 0.610 & 1 \\
& & & & 15355 & 250.8$\pm$15 & 36 & \SIII $\lambda$9532 & -- & -- & 1 \\
186 & 53.0555878 & -27.7234249 & 23.23 & 11436 & 129.1$\pm$38 & 189 & \OIII\ & 16.5$\pm$4 & 1.294 & 1 \\
& & & & 15056 & 54.7$\pm$15 & 81 & \Ha\ & 4.3$\pm$1 & 1.294 & 1 \\
262 & 53.0558815 & -27.7187843 & 25.63 & 13650 & 47.8$\pm$14 & 248 & \OIII\ & 12.6$\pm$3 & 1.733 & 3 \\
226\tablenotemark{\ddag} & 53.0558815 & -27.7210884 & 24.04 & 13406 & 35.6$\pm$11 & 100 & \Ha\ & 1.6$\pm$0 & 1.043 & 1 \\
& & & & 10229 & 99.5$\pm$31 & 110 & \OIII\ & 7.6$\pm$2 & 1.043 & 1 \\
316 & 53.0565033 & -27.7156715 & 23.33 & 13254 & 115.4$\pm$12 & 125 & \Ha\ & 5.0$\pm$0 & 1.020 & 2 \\
238 & 53.0569191 & -27.7202835 & 22.85 & 12453 & -- & -- & \OIII\ & -- & 1.493 & 1 \\
& & & & 16247 & -- & -- & \Ha\ & -- & 1.476 & 1 \\
336\tablenotemark{\dag}  & 53.0577126 & -27.7135887 & 20.71 & 8682 & 259.9$\pm$48 & 59 & \OIII\ & 8.4$\pm$1 & 0.738 & 1 \\
& & & & 15745 & 32.8$\pm$11 & 16 & \SIII $\lambda$9069 & -- & -- & 1 \\
& & & & 11699 & 69.1$\pm$18 & 35 & \SII & -- & -- & 1 \\
& & & & 11430 & 222.4$\pm$21 & 62 & \Ha\ & 4.4$\pm$0 & 0.738 & 1 \\
& & & & 16549 & 53.3$\pm$11 & 28 & \SIII $\lambda$9532 & -- & -- & 1 \\
411 & 53.0586700 & -27.7082272 & 25.84 & 13352 & 49.6$\pm$13 & 388 & -- & -- & -- & -- \\
578 & 53.0589218 & -27.6978111 & 24.33 & 16636 & 104.9$\pm$18 & 133 & \OIII\ & 57.2$\pm$9 & 2.315 & 1 \\
& & & & 12354 & 49.6$\pm$16 & 67 & \OII\ & 28.6$\pm$9 & 2.315 & 1 \\
559 & 53.0613785 & -27.6981163 & 22.15 & 9317 & 47.8$\pm$21 & 21 & \Ha\ & 0.2$\pm$0 & 0.420 & 1 \\
539 & 53.0624619 & -27.6987267 & 25.04 & 15709 & 39.6$\pm$12 & 212 & -- & -- & -- & -- \\
103 & 53.0633392 & -27.7272835 & 26.87 & 13397 & 61.0$\pm$17 & 121 & \OIII\ & 15.0$\pm$4 & 1.682 & 1 \\
& & & & 9987 & 39.3$\pm$15 & 73 & \OII\ & 10.4$\pm$3 & 1.682 & 1 \\
427 & 53.0643387 & -27.7056999 & 22.48 & 9294 & 48.4$\pm$21 & 31 & \Ha\ & 0.2$\pm$0 & 0.416 & 1 \\
195 & 53.0656700 & -27.7203941 & 23.34 & 13711 & 94.0$\pm$15 & 103 & \OIII\ & 25.3$\pm$4 & 1.745 & 1 \\
& & & & 10224 & 53.7$\pm$15 & 19 & \OII\ & 15.5$\pm$4 & 1.745 & 1 \\
364 & 53.0693359 & -27.7090893 & 23.52 & 10775 & 102.6$\pm$14 & 73 & \Ha\ & 1.4$\pm$0 & 0.642 & 1 \\
& & & & 8219 & 376.6$\pm$98 & 270 & \OIII\ & 8.7$\pm$2 & 0.642 & 1 \\
246 & 53.0700035 & -27.7165890 & 24.95 & 8489 & 70.1$\pm$34 & 72 & \OIII\ & 2.0$\pm$0 & 0.696 & 1 \\
& & & & 11133 & 48.7$\pm$23 & 273 & \Ha\ & 0.8$\pm$0 & 0.696 & 1 \\
215 & 53.0703392 & -27.7178669 & 23.46 & 10935 & -- & -- & \Ha\ & -- & 0.666 & 1 \\
563 & 53.0705452 & -27.6956444 & 23.77 & 9949 & 53.8$\pm$17 & 36 & \OII\ & 14.0$\pm$4 & 1.673 & 1 \\
& & & & 13353 & 93.1$\pm$14 & 143 & \OIII\ & 22.6$\pm$3 & 1.673 & 1 \\
402 & 53.0712967 & -27.7058105 & 24.94 & 13730 & 133.0$\pm$10 & 656 & \OIII\ & 35.9$\pm$2 & 1.749 & 3 \\
211\tablenotemark{\dag}  & 53.0714226 & -27.7175884 & 20.41 & 10267 & 478.8$\pm$28 & 44 & \Ha\ & 4.9$\pm$0 & 0.564 & 1 \\
476 & 53.0715446 & -27.7006989 & 25.27 & 16555 & 67.8$\pm$10 & 353 & -- & -- & -- & -- \\
193 & 53.0723381 & -27.7186718 & 20.98 & 10813 & 119.6$\pm$25 & 27 & \Ha\ & 1.7$\pm$0 & 0.648 & 1 \\
175 & 53.0725441 & -27.7198391 & 24.38 & 12984 & 37.2$\pm$10 & 231 & \Ha\ & 1.5$\pm$0 & 0.978 & 3 \\
583 & 53.0730896 & -27.6939487 & 24.17 & 15239 & 41.7$\pm$12 & 225 & \Ha\ & 3.4$\pm$1 & 1.322 & 2 \\
250 & 53.0734215 & -27.7159519 & 22.03 & 10819 & 227.1$\pm$18 & 86 & \Ha\ & 3.2$\pm$0 & 0.649 & 2 \\
210 & 53.0735054 & -27.7173939 & 22.71 & 10845 & 90.0$\pm$16 & 34 & \Ha\ & 1.3$\pm$0 & 0.653 & 1 \\
445 & 53.0736694 & -27.7024498 & 23.22 & 14368 & 67.9$\pm$11 & 68 & \Ha\ & 4.3$\pm$0 & 1.189 & 2 \\
566 & 53.0740471 & -27.6945629 & 22.80 & 15303 & 112.4$\pm$13 & 73 & \Ha\ & 9.4$\pm$1 & 1.332 & 2 \\
474 & 53.0740891 & -27.7001171 & 24.31 & 13713 & 99.4$\pm$12 & 121 & \OIII\ & 26.7$\pm$3 & 1.746 & 1 \\
454 & 53.0761719 & -27.7011452 & 22.76 & 9264 & 59.2$\pm$16 & 40 & \OIII\ & 2.7$\pm$0 & 0.852 & 1 \\
& & & & 12157 & 86.1$\pm$14 & 88 & \Ha\ & 2.4$\pm$0 & 0.852 & 1 \\
251 & 53.0767975 & -27.7144222 & 25.20 & 14213 & 92.9$\pm$13 & 162 & \OIII\ & 28.6$\pm$4 & 1.846 & 3 \\
397 & 53.0772133 & -27.7047558 & 26.86 & 13821 & 96.9$\pm$13 & 254 & \OIII\ & 26.9$\pm$3 & 1.760 & 1 \\
263 & 53.0772972 & -27.7131157 & 23.88 & 14569 & 39.8$\pm$12 & 82 & \Ha\ & 2.7$\pm$0 & 1.222 & 1 \\
& & & & 11097 & 27.0$\pm$12 & 24 & \OIII\ & 3.0$\pm$1 & 1.222 & 1 \\
339 & 53.0773392 & -27.7081985 & 22.65 & 8002 & 863.9$\pm$163 & 375 & \OIII\ & 16.9$\pm$3 & 0.602 & 1 \\
& & & & 10512 & 145.8$\pm$17 & 83 & \Ha\ & 1.7$\pm$0 & 0.602 & 1 \\
265 & 53.0780029 & -27.7129498 & 25.34 & 13775 & 42.5$\pm$12 & 182 & -- & -- & -- & -- \\
351 & 53.0785751 & -27.7074108 & 22.62 & 13744 & 92.2$\pm$25 & 97 & \Ha\ & 4.8$\pm$1 & 1.094 & 2 \\
498 & 53.0789223 & -27.6977272 & 21.52 & 12996 & 181.0$\pm$20 & 77 & \Ha\ & 7.1$\pm$0 & 0.980 & 3 \\
512 & 53.0792542 & -27.6968098 & 25.15 & 13097 & 52.0$\pm$11 & 301 & \OIII\ & 11.7$\pm$2 & 1.622 & 1 \\
10 & 53.0816612 & -27.6881046 & 22.72 & 15232 & 79.5$\pm$30 & 59 & \Ha\ & 6.5$\pm$2 & 1.321 & 3 \\
242 & 53.0821304 & -27.7137547 & 24.61 & 11506 & 55.3$\pm$21 & 97 & \OII\ & 24.8$\pm$9 & 2.070 & 1 \\
& & & & 15337 & 62.9$\pm$14 & 111 & \OIII\ & 25.7$\pm$5 & 2.070 & 1 \\
3\tablenotemark{\dag} & 53.0825462 & -27.6896439 & 18.67 & 8054 & 858.4$\pm$203 & 40 & \Ha\ & 1.0$\pm$0 & 0.227 & 1 \\
418 & 53.0848389 & -27.7014771 & 24.92 & 14051 & 182.6$\pm$31 & 344 & \OIII\ & 53.9$\pm$9 & 1.813 & 3 \\
437 & 53.0855865 & -27.6999226 & 24.70 & 14053 & 88.6$\pm$26 & 501 & \OIII\ & 26.2$\pm$7 & 1.814 & 3 \\
258 & 53.0857124 & -27.7113400 & 24.32 & 9998 & 115.9$\pm$31 & 65 & \OIII\ & 7.9$\pm$2 & 0.998 & 1 \\
& & & & 13111 & 38.4$\pm$10 & 99 & \Ha\ & 1.6$\pm$0 & 0.998 & 1 \\
416 & 53.0860062 & -27.7011986 & 24.26 & 12760 & 163.7$\pm$34 & 567 & -- & -- & -- & -- \\
146 & 53.0872955 & -27.7184486 & 24.44 & 11539 & 50.2$\pm$24 & 159 & \OIII\ & 6.7$\pm$3 & 1.309 & 1 \\
& & & & 15156 & 46.1$\pm$11 & 138 & \Ha\ & 3.7$\pm$0 & 1.309 & 1 \\

\enddata
\tablenotetext{*}{NOTE: No data in the case of fluxes (and EW, SFR) indicates that the spectrum had a high level of contamination but wavelengths were secure enough to warrent redshift determination.  In the case of line IDs, no data indicates that no suitable line ID was found for the given input redshift.  ``Grism Redshift'' column gives re--calculated redshift based on the line identification.  ``Flag'' column gives source of input redshift used for line identification, where used: 1---two lines visible in spectrum, no prior redshift needed; 2---single line in spectrum, line ID and grism redshift based on prior spectroscopic redshift; 3---single line in spectrum, line ID and grism redshift based on prior photometric redshift.}
\tablenotetext{\dag}{CDF-S X-ray sources identified as AGN by Szokoly \etal 2004.}
\tablenotetext{\ddag}{New spectroscopic redshift (no previous photometric or spectroscopic redshift measurements).}
\end{deluxetable}


\begin{thebibliography}{}
%\input bibliography_wfc3ers.tex
\bibitem[atex]{atex} Atek, H. \etal 2010, ApJ, 723, 104
\bibitem[bal]{bal} Balestra, I. \etal 2010, A\&A, 512, A12
\bibitem[bau]{bau} Bauer, A.E., Drory, N., \& Feulner, G. 2005, ApJL, 621, 89
\bibitem[beck]{beck} Beckwith, S.V.W. \etal 2006, AJ, 132, 1729
\bibitem[Bertin \& Arnouts(1996)]{sex} Bertin, E. \& Arnouts, S. 1996, A\&AS, 117, 363
\bibitem[bc03]{bc03} Bruzual, G. \& Charlot, S. 2003, MNRAS 344, 1000
\bibitem[bou]{bou} Bouwens, R.J. \etal 2010a, ApJL, 709
\bibitem[bou2]{bou2} ----- 2010b, ApJL, 708, 69
\bibitem[dam]{dam} Damen, M., Labbé, I., Franx, M., van Dokkum, P. G., Taylor, E. N., Gawiser, E. J. 2009, ApJ, 690, 937
\bibitem[md]{md} Dickinson, M., et al. 2003, Proc. ESO/USM Workshop, ``The Mass in Galaxies at Low and High Redshift'', ed. R. Bender, \& A. Renzini (Berlin: Springer), 324
\bibitem[faz]{faz} Fazio, G. G., et al. 2004, ApJS, 154, 10
\bibitem[feu]{feu} Feulner, G., Goranova, Y., Drory, N., Hopp, U., \& Bender, R. 2005, MNRAS, 358, L1
\bibitem[fkl]{fkl} Finkelstein, S.L. \etal 2010, ApJ, 719, 1250
\bibitem[gal]{gal} Gallego, J., Zamorano, J., Aragon-Salamanca, A., \& Rego,M. 1995, ApJ 455, L1
\bibitem[gard]{gard} Gardner, J.P., \etal 2006, SSR, 123, 485
\bibitem[gia]{gia} Giavalisco, M. \etal 2004, ApJL, 600, 93
\bibitem[graz]{graz} Grazian, A., \etal 2006, A\&A 449, 951
\bibitem[grog]{grog} Grogin, N.A., Malhotra, S., Rhoads, J., Cohen, S., Hathi, N., Windhorst, R., Pirzkal, N. 2007, BAAS, 211, 4605
\bibitem[nph]{nph} Hathi, N.P., \etal 2010, ApJ, 720, 1708
\bibitem[k83]{k83} Kennicutt, R.C.,Jr. 1983, ApJ 272, 54
\bibitem[k92]{k92} ----- 1992, ApJ, 388, 310
\bibitem[k98]{k98} ----- 1998, ARA\&A 36, 189
\bibitem[k00]{k00} Kennicutt, R.C., Jr., Bresolin, F., French, H., \& Martin, P. 2000, ApJ, 537, 589
\bibitem[kewl]{kewl} Kewley, L.J., Geller, M.J., Jansen, R.A. 2004, AJ, 127, 2002
\bibitem[koe]{koe} Koekemoer, A. M., Fruchter, A. S., Hook, R. N., \& Hack, W. 2002, The 2002 HST Calibration Workshop, ed. S. Arribas, A. Koekemoer, and B. Whitmore (Baltimore:STScI), 337
\bibitem[kue]{kue} K{\"u}mmel, M., Walsh, J.~R., Pirzkal, N., Kuntschner, H., \& Pasquali, A.\ 2009, PASP, 121, 59
\bibitem[kun]{kun} H. Kuntschner, H. Bushouse, M. K{\"u}mmel, J. R. Walsh 2009a, ST-ECF ISR WFC3-2009-18 WFC3 SMOV proposal 11552: Calibration of the G102 grism
\bibitem[kb]{kb} ----- 2009b, ST-ECF ISR WFC3-2009-17 WFC3 SMOV proposal 11552: Calibration of the G141 grism
\bibitem[lb]{lb} Labbe, I. \etal 2010, ApJL, 716, 103
\bibitem[lt78]{lt78} Larson, R.B. \& Tinsley, B.M. 1978, ApJ 219, 46
\bibitem[li08]{li08} Li \etal 2008, MNRAS,385, 1903L
\bibitem[mart]{mart} Martin, D. C., et al. 2007, ApJS, 173, 415
\bibitem[mcc]{mcc} McCarthy, P.J., \etal 1999, ApJ, 520, 548
\bibitem[mclure]{mclure} McLure, R. J., Dunlop, J. S., Cirasuolo, M., Koekemoer, A. M., Sabbi, E., Stark, D. P., Targett, T. A., Ellis, R. S. 2010, MNRAS, 403, 960
\bibitem[gm]{gm} Meurer \etal 2007, AJ, 134, 77
\bibitem[nav97]{nav97} Navarro, J., Frenk, C., \& White, S. 1997, ApJ, 490, 493
\bibitem[noe]{noe} Noeske, K. \etal 2007, ApJL, 660, 43 
\bibitem[oesch]{oesch} Oesch, P. A., Bouwens, R. J., Carollo, C. M., Illingworth, G. D., Trenti, M., Stiavelli, M., Magee, D., Labbé, I., Franx, M. 2010, ApJL, 709, 21
\bibitem[ovzr]{ovzr} Overzier, R.A. \etal 2008, ApJ, 677, 37
\bibitem[pergon]{pergon} Perez-Gonzalez, P. G., et al. 2008, ApJ, 675, 234
\bibitem[pg06]{pg06} ----- 2006, ApJ 636, 582
\bibitem[rod]{rod} Rodighiero, G. \etal 2010, A\&A, 518, L25
\bibitem[rosa]{rosa} Rosa, M., Joubert, M., Benvenuti, P. 1984, A\&AS, 57, 361
\bibitem[me08]{me08} Straughn, A.N. et al. 2008, AJ, 135, 1624
\bibitem[me09]{me09} Straughn, A.N. et al. 2009, AJ, 138, 1022
\bibitem[sz09]{sz09} Szokoly, G.P. \etal 2004, ApJS, 155, 271
\bibitem[vd10]{vd10} van Dokkum, P.G \& Brammer, G. 2010, arXiv:1003:3446
\bibitem[wf91]{wf91} White, S. D. M., \& Frenk, C. S. 1991, ApJ, 379, 52
\bibitem[wil10]{wil10} Wilkins, S. M., Bunker, A. J., Ellis, R. S., Stark, D., Stanway, E. R., Chiu, K., Lorenzoni, S., Jarvis, M. J. 2010, MNRAS, 403, 938
\bibitem[raw10]{raw10} Windhorst, R.A. et al. 2010, submitted, arXiv:1005.2776
\bibitem[van]{van} Vanzella, E. \etal 2008, A\&A, 478, 83V
\bibitem[wuyts]{wuyts} Wuyts, S., Labb´e, I., Schreiber, N. M. F., et al. 2008, ApJ, 682, 985
\bibitem[xia]{xia} Xia, L. \etal 2010, submitted 
\bibitem[xu]{xu} Xu, C., \etal 2007, AJ 134, 169
\bibitem[haojing]{haojing}  Yan, H. Windhorst, R.A., Hathi, N.P., Cohen, S.H., Ryan, R.E., O'Connell, R., McCarthy, P.J. 2010a, RAA, 10, 867
\bibitem[haojing2]{haojing2} Yan, H. \etal 2010b, ApJ, submitted 
\bibitem[yan]{yan} Yan, L., McCarthy, P.J., Freudling, W., Teplitz, H.I., Malumuth, E.M., Weymann, R.J., \& Malkan, M.A. 1999, ApJL 519, L47
\bibitem[zheng]{zheng} Zheng, X. Z., Bell, E. F., Papovich, C.,Wolf, C., Meisenheimer, K., Rix, H.-W.,
Rieke, G, \& Somerville, R. 2007, ApJ, 661, 41
\end{thebibliography}
\end{document}